\pdfoutput=1
\documentclass[onecolumn]{sdl}

\def\0{\mbox{\tiny $0$}}
\def\1{\mbox{\tiny $1$}}
\def\2{\mbox{\tiny $2$}}
\def\3{\mbox{\tiny $3$}}
\def\4{\mbox{\tiny $4$}}
\def\5{\mbox{\tiny $5$}}
\def\6{\mbox{\tiny $6$}}
\def\7{\mbox{\tiny $7$}}
\def\8{\mbox{\tiny $8$}}
\def\9{\mbox{\tiny $9$}}

\logo{
\colorbox{DarkGoldenrod}{\color{white}$\mathbf{\Sigma\hspace*{0.06cm} \delta \hspace*{0.04cm} \Lambda}$}
}

\journal{\shadowtext{\textbf{\color{DarkRed} Journal of Modern Optics}} \,\,{\textbf 65}, 837-846 (2018).}

\titlelines{2}
\title{Optical weak measurements without\\ removing the Goos-H\"{a}nchen phase}

\imgbgabstract{Optical Weak Measurements are a powerful tool for measuring small shifts of optical paths. When applied to the measurement of the Goos-H\"{a}nchen shift, in particular, a special step must be added to its protocol: the removal of the relative Goos-H\"anchen phase, since its presence generates a destructive influence on the measurement. There is, however, a lack of description in the literature of the precise effect of the Goos-H\"{a}nchen phase on weak measurements. In this paper we address this issue, developing an analytic study for a Gaussian beam transmitted through a dielectric structure. We obtain analytic expressions for weak measurements as a function of the relative Goos-H\"{a}nchen phase and show how to remove it without the aid of waveplates.
}

\author{
\names{Manoel P. Ara\'ujo$^{1,a}$, Stefano De Leo$^{2,b}$ and Gabriel G. Maia$^{1,c}$}
\affiliation{$^{1}$Institute of Physics Gleb Wataghin, State University of Campinas, Brazil.}
\email{$^{a}$mparaujo@ifi.unicamp.br $\,\,\,\,\, $ $^{c}$ggm11@ifi.unicamp.br}
\affiliation{$^{2}$Department of Applied Mathematics, State of University of Campinas, Brazil.}
\email{$^{b}$deleo@ime.unicamp.br}
}

\begin{document}

\sdlmaketitle

\section*{\shadowtext{ I. Introduction}}

%
%
%
%
%
%
%
%

The geometrical Optics is well suited for plane waves \cite{B1999,B2007}, but for bounded light beams corrections must be made to the optical path when interactions with interfaces between different media take place. In particular, one of the better known of such corrections is the so-called Goos-H\"anchen shift\cite{GH1947,GH1948,GH1949,GH1971,GH1971p1397,GH1988}, which induces spatial translations of the optical path in the plane of incidence for totally internally reflected beams. It was first observed experimentally by Herman Goos and Hilda H\"{a}nchen in 1947 \cite{GH1947} for transverse electric (TE) polarized light, being theoretically analyzed by Artmann the following year \cite{GH1948}. Artmann also presented in his paper an analysis of the transverse magnetic (TM) polarization, which was confirmed experimentally in 1949 by Goos and H\"{a}nchen \cite{GH1949}. This phenomenon is present in several different optical systems such as waveguides\cite{GH2007}, photonic crystals\cite{GH2004}, and resonators\cite{GH2006}, but because it is intrinsically linked to the wave nature of the electromagnetic field, it can also be observed in other oscillatory systems such as for acoustic \cite{GH2000} and seismic waves\cite{GH2012}.

The measurement of this shift is usually a challenging task because it is a minute phenomenon when compared to the transverse extent of the beam, which is typically of a few wavelengths. To deal with this problem, optical Weak Measuments have been successfully employed \cite{S2013,CR2015}. This technique is an optical analog of the quantum Weak Measurement, which was devised originally for spin-$1/2$ particles \cite{S1988,S1989}, playing the TE and TM polarization modes the roles of the \emph{up} and \emph{down} modes. The principle behind it is simple. A laser beam diagonally polarized is made to interact with a dielectric structure, which will produce a different shift for each of its components. A polarizer then mixes these components producing an intensity profile which is dependent on the relative shift between the TE and TM contributions. By changing the polarization angle we change the position of the intensity peaks and by measuring the distance between such peaks it is possible to indirectly measure the relative shift. Under appropriate conditions the distance between the peaks is greater than the Goos-H\"{a}nchen shift, hence the status of weak measurements as an amplification technique.


An important step of this process is the phase removal. Upon interaction with the dielectric the light beam is not only shifted, but also acquires an additional polarization-dependent phase, the Goos-H\"{a}nchen phase. In Weak Measurement experiments the relative phase between polarization states is usually removed with the aid of waveplates before the second polarizer \cite{S2013}, since it creates an interference effect which is destructive to the measurement. To the best of our knowledge, however, up to this point, there is lack of a formal description of the effect of the Goos-H\"{a}nchen phase in optical weak measurements, which is the point we hope to address in this paper. Since this phase is a function of the incidence angle and of the number of reflections inside the dielectric, it is possible to control it by changing this angle as well as the length of the prism. For particular choices of these parameters the Goos-H\"{a}nchen phase can be completely removed, without need of waveplates. This analysis is also relevant as an efficiency test for such devices. Since the relative phase has a destructive effect on measurements, a waveplate that leaves a residual phase may compromise experimental results. Knowing how phases affect weak measurements may helps us identify when they were not completely removed.

The paper is organized as follows: in section II we describe the optical system used, the electric fields involved as well as the Goos-H\"{a}nchen phase induced by the Fresnel coefficients of the prism. Section III is dedicated to obtain the closed expression for weak measurement with phase. We show how  our result reproduce the usual formula found in the literature. In section IV we discuss these results, analyzing the influence of the Goos-H\"{a}nchen phase on measurements. Finally, in section V, we present our conclusions and future perspectives.

\section*{\shadowtext{II. The optical system}}

The optical system of interest is shown in Fig.1(a). The laser source generates an unpolarized Gaussian beam which then passes through the first polarizer, set at an angle of $\pi/4$, making it an equal mixture of TE and TM polarized waves. After this, it enters the dielectric structure from its left interface, making an angle $\theta$ with its normal and being refracted with an angle $\psi$. Both angles are connected via the Snell's law, $\sin\theta=n\,\sin\psi$, where $n$ is the refractive index of the structure. Once inside it, the beam undergoes multiple total internal reflections at the down and upper interfaces of the dielectric. The incidence (and reflection) angle at these interfaces is given by $\varphi=\psi+\pi/4$. After exiting the prism, light passes through the second polarizer, whose polarization angle is a controllable parameter of the system, being subsequently collected by the camera. The dielectric structure is built as a chain of an even number of right angle triangular prisms, as shown in Fig.1(b). Since each triangular prism amounts to one internal reflection, the total number of reflections can be controlled by the number of prisms in the chain. The even number allows the transmitted beam to be parallel to the incident one. This choice is made to simplify the geometrical description of the system.

Let us start our analysis by considering the Guassian light beam propagating from the source to the dielectric.
After the first polarizer have selected it into a mixed polarization state, assuming polarization in the $x$ (TE) and $y$ (TM) directions and propagation in the $z$ direction, the electric field is given by
\begin{equation}
\boldsymbol{E}_{_{\mathrm{INC}}}(x,y,z)=E(x,y,z)\left(\boldsymbol{\hat{x}}+\boldsymbol{\hat{y}}\right),
\end{equation}
with
\begin{equation}
\label{eq:Ein}
E(x,y,z) =  \frac{\displaystyle E_{\0}\,\,\,e^{ik\,z}}{\displaystyle 1+i\,z/z_{_{R}}} \exp \left[-\frac{\displaystyle x^{\2}+y^{\2}}{\displaystyle {\mathrm{w}}_{\0}^{\2}\,\left(1+i\,z/z_{_{R}}\right)}\right]\,,
\end{equation}
where
 $
  z_{_{R}} = k\, \mathrm{w}_{\0}^{\2}/2
$
is the Rayleigh distance, $k=2\pi/\lambda$ the wave number, being $\lambda$ the wavelength, $\mathrm{w}_{\0}$ the minimum beam waist, and $\boldsymbol{\hat{x}}$ and $\boldsymbol{\hat{y}}$ unit vectors. The incoming intensity profile is
\begin{equation}
\mathcal{I}_{_0}(x,\,y,\,z) = E_{\0}^{^2}\,\frac{\displaystyle \mathrm{w}^{\2}_{\0}}{\displaystyle \mathrm{w}^{\2}(z)}\,\exp\left[-2\frac{\displaystyle \,x^{\2}+y^{\2}}{\displaystyle \mathrm{w}^{\2}(z)}\right],
\end{equation}
where
\begin{equation}
\mathrm{w}(z) =  \mathrm{w}_{\0} \sqrt{ 1 + \left( z/ z_{_{R}} \right)^{\2}}.
\end{equation}

The beam then hits the left interface of the dielectric structure at an angle compatible with the total internal reflection regime, which happens for

\[\varphi\geq\arcsin\left[\frac{1}{n}\right],\]
implying the following relation for the incidence angle,

\begin{equation}
\theta\geq\arcsin\left[\frac{1-\sqrt{n^{\2}-1}}{\sqrt{2}}\right].
\end{equation}
For a borosilicate glass prism, $n=1.515$, we have that $\theta\geq -5.609^{\circ}$.
\\

Inside the prism, multiple internal reflections take place and the outputted beam is modified by the Fresnel transmission coefficient of the dielectric, which is given by the product of the Fresnel coefficients at each interface:
\begin{equation}{\label{eq:TE}}
T_{\sigma}= \frac{4 \, a_{\sigma} \, \cos\theta \, \cos\psi}{(a_{\sigma} \, \cos\theta +\, \cos\psi)^{\2}}\, \exp[i\,\left(\,\Phi_{\0}+\Phi_{\sigma}\,\right)],
\end{equation}
where
\[a_{\sigma} = \{a_{_\mathrm{TE}},\,\,a_{_\mathrm{TM}} \}\,\,=\,\,\left\{1/n,\,n\right\}.\,\,\]
$\Phi_{\0}$ is the geometrical phase, which depends on the geometry of the system, as the name suggests,
 and is independent of the polarization state. It is given by
\begin{equation}
 \Phi_{\0} =N k\left(\sqrt{2}\,n\,\cos\varphi+ n\cos\psi-\cos\theta\,\right) b\,,
 \end{equation}
being the parameter $b$ the small length of the planar section of the structure, see Fig.1(b), and $N$ the number of internal reflections. The Goos-H\"anchen phase, which is independent of the geometry of the system and dependent on the polarization state of the beam, is given by
 \begin{equation}
\Phi_{\sigma} = -2\,N \,\arctan\left[ a_{\sigma} \frac{\displaystyle \sqrt{n^{\2}\,\sin^{\2}\varphi-1}}{\displaystyle \cos\varphi} \right]\,\,.
\end{equation}

The optical path and the radiation's phase are intrinsically related and by using the Stationary Phase Method \cite{SPM2015} the first can be calculated by a first order derivative with respect to incidence angle of the second. Applying this method to the geometrical phase we obtain, in the direction orthogonal to the propagation direction of the beam, the distance between the incoming and the outgoing beams, which are parallel,
\begin{eqnarray}
y_{\0}=\Phi_{\0}^{'}/k = N\,\Big{(}\,\cos\theta-\sin\theta+2\,\tan\psi\,\cos\theta\,\Big{)}\,b\,\,,
\end{eqnarray}
which is of the order of the length of the block. Applying this method to the Goos-H\"anchen phase we find the Goos-H\"anchen shifts,
\begin{equation}
 y_{_\mathrm{TE}}=\Phi_{_\mathrm{TE}}^{'}/k=\frac{\displaystyle N\,\cos\theta\,\sin\varphi}{\displaystyle \pi\,\cos\psi\,\sqrt{n^{\2}\sin^{\2}\varphi-1}}\,\lambda
\end{equation}
and
\begin{equation}
 y_{_\mathrm{TM}} = \Phi_{_\mathrm{TM}}^{'}/k = \frac{\displaystyle y_{_\mathrm{TE}}}{\displaystyle n^{\2}\sin^{\2}\varphi-\cos^{\2}\varphi}
\end{equation}
for TE and TM polarized light, respectively.
\\

The transmitted beam is then given by
\begin{equation}
\boldsymbol{E}_{_{\mathrm{TRA}}}(x,y,z)=E_{_{\mathrm{TE}}}(x,y,z)\boldsymbol{\hat{x}}+E_{_{\mathrm{TM}}}(x,y,z)\boldsymbol{\hat{y}}
\end{equation}
where the approximation
\begin{eqnarray}{\label{eq:Et}}
E_{\sigma}(x,\,y,\,z) \, &\approx& \, \left|T_{\sigma}\right| \, E\left(x,\, y - y_{\0} - y_{\sigma},\,z \right)\,\exp\left[i \left( \Phi_{\0}+\Phi_{\sigma} \right)\right]\,\,,
\end{eqnarray}
is used. By passing it through the second polarizer, at an angle $\beta$ with respect to the $x$-axis, the electric field amplitude becomes the weighted sum of the components $E_{\sigma}$, having then the beam collected by the camera the form
\begin{equation}
\boldsymbol{E}_{_{\mathrm{CAM}}}(x,y,z)=\left[\,\cos\beta\,E_{_{\mathrm{TE}}}(x,y,z)+\sin\beta\,E_{_{\mathrm{TM}}}(x,y,z)\,\right]\,\left(\cos\beta\,\boldsymbol{\hat{x}}+\sin\beta\,\boldsymbol{\hat{y}}\right)\,\,,
\end{equation}
being the associated intensity profile
\begin{equation}{\label{eq:Iw}}
\mathcal{I} = \left|\cos\beta \, E_{_\mathrm{TE}}(x,\,y,\,z) + \sin\beta \, E_{_\mathrm{TM}}(x,\,y,\,z) \right|^{\2}\,\,.
\end{equation}
Notice that the intensity is the superposition of two Gaussian functions, with displaced centers and a relative phase between them. This generates a double peaked curve, with the peaks controlled by the weights of the Gaussians, the sine and cosine functions. In the next section we will evaluate the position of the peaks as a way to measure the relative Goos-H\"{a}nchen shift.

\section*{\shadowtext{III. The optical weak measurement}}

By putting Eq.(\ref{eq:Et}) in Eq.(\ref{eq:Iw}), and with the aid of the definitions,
\begin{eqnarray}
\begin{tabular}{c}
$Y \, = \, \frac{\displaystyle 1}{\displaystyle \mathrm{w}(z)}\left(y - y_{\0} - \frac{\displaystyle y_{_{\mathrm{TE}}} + y_{_{\mathrm{TM}}} }{\displaystyle 2}\right)$
\\\\
$\Delta y_{_{\mathrm{GH}}} \, = \, \frac{\displaystyle y_{_{\mathrm{TM}}} - y_{_{\mathrm{TE}}}}{\displaystyle \mathrm{w}(z)}$
\\\\
$\tau \, = \, \left| T_{_{\mathrm{TE}}} \right|/\left| T_{_{\mathrm{TM}}} \right|$
\\\\
$\Delta\Phi_{_{\mathrm{GH}}}= \Phi_{_{\mathrm{TE}}}-\Phi_{_{\mathrm{TM}}} = 2 \, N \arctan\left[\frac{\displaystyle \sqrt{n^{\2} \, \sin^{\2}\varphi-1}}{\displaystyle n \, \sin\varphi\,\tan\varphi}\right] \,\,\,,$
\end{tabular}
\end{eqnarray}
we obtain the outputted intensity
\begin{eqnarray}{\label{eq:Iwa}}
\mathcal{I}  &\propto& \bigg{|}  \tau\,\exp\left[ - \left( Y + \Delta y_{_{\mathrm{GH}}}/2 \right)^{\2}\right]\exp\left[i\, \Delta\Phi_{_{\mathrm{GH}}}\right]+ \tan\beta \,\exp\left[ -\left( Y - \Delta y_{_{\mathrm{GH}}}/2 \right)^{\2} \right] \bigg{|}^{\2}\,\,.
\end{eqnarray}
A few approximations can be made to Eq.(\ref{eq:Iwa}). The difference in the transmission of TE and TM waves can be neglected, rendering $\tau \approx 1$.  Besides that, by choosing the angle of the second polarizer as
\begin{equation*}
\beta\,=\,3\pi/ 4+\!\Delta\epsilon\,\,,
\end{equation*}
and considering a very small perturbation $\Delta\epsilon$ about the fixed angle $3\pi/ 4$, that is, $\Delta\epsilon \ll 1$,  it can be shown that
\begin{equation}
\tan\beta \, \approx \, 2\Delta\epsilon-1\,\,.
\end{equation}
With these approximations, the electric field intensity collected by the camera is then
\begin{eqnarray}
\mathcal{I} &\propto& \bigg{|}\,\left(1 -Y\,\Delta y_{_{\mathrm{GH}}} \right) \exp\left[i\,\Delta\Phi_{_{\mathrm{GH}}}\right]+ (2\Delta\epsilon-1)\,\left(1 + Y\,\Delta y_{_{\mathrm{GH}}} \right) \bigg{|}^{\2} \,  \mathrm{exp}\left[-2\,Y^{\2} \right]\,\,,
\end{eqnarray}
which can be simplified as
\begin{eqnarray}
\mathcal{I} &\propto&\left[ \left( \Delta\epsilon -  Y\, \Delta y_{_{\mathrm{GH}}} \right)^{\2} + \sin^{\2}\left( \Delta\phi_{_{\mathrm{GH}}}/ 2\right)\right]\,\mathrm{exp}\left[-2\,Y^{\2} \right]\,\,.
\end{eqnarray}
We can see that this equation is a function of two significant physical quantities, the Goos-H\"anchen shift and phase, and one experimental parameter, the perturbation rotation angle. The positions of the intensity maximums are given by the solutions of the quadratic equation
\begin{equation}
 Y^{\2}-2\,A\,Y-1/4=0\,\,,
\end{equation}
which are
\begin{equation}
 Y_{_{\mathrm{MAX}}}^{\pm} =A\,\pm\,\sqrt{A^{\2}+\,\,1/4}\,\,,
\end{equation}
with
\begin{equation*}
A=\left[2\,\Delta\epsilon^{\2}-\Delta y^{\2}_{_{\mathrm{GH}}}+2\,\sin^{\2}\left(\Delta\phi_{_{\mathrm{GH}}}/ 2\right)\right]\Big{/}\,8\,\Delta\epsilon\, \Delta y_{_{\mathrm{GH}}}\,\,.
\end{equation*}
Only the positive solution will be considered, since the intensity associated to it is the greater one. By making two successive measurements of the intensity, one for a counterclockwise rotation ($-|\Delta\epsilon|$) of the second polarizer and one for a clockwise rotation ($|\Delta\epsilon|$), the intensity peak is displaced, and the distance between both positions,
\begin{eqnarray}
\Delta Y_{_{\mathrm{MAX}}} &=& Y_{_{\mathrm{MAX}}}^{+}(-|\Delta\epsilon|)-Y_{_{\mathrm{MAX}}}^{+}(|\Delta\epsilon|)\,\,,
\end{eqnarray}
can be used to indirectly determine the relative Goos-H\"{a}nchen shift. Remembering that the shift is of the order of the wavelength and choosing $\Delta\epsilon$ to satisfy the condition
\[ |\Delta y_{_{\mathrm{GH}}}| \, \ll \, \sqrt{2}\,|\Delta\epsilon|\,\,, \]
we obtain
\begin{equation}{\label{eq:Ymax}}
\Delta Y_{_{\mathrm{MAX}}} \,=\, \frac{\displaystyle \Delta y_{_{\mathrm{GH}}}}{\displaystyle |\Delta\epsilon|}\left\{ 1+\left[\frac{\displaystyle \sin(\Delta\Phi_{_{\mathrm{GH}}}/2)}{\displaystyle |\Delta\epsilon|}\right]^{\2}\right\}^{-1}
\,\,.
\end{equation}
This equation represents the standard Weak Measurement formula modified by the Goos-H\"anchen phase. A detailed analysis of its effect on measurements is provided in the next section.

\section*{\shadowtext{ IV. The Goos-H\"anchen phase} }

In the laboratory, one measures the distance between intensity maximums, $\Delta Y_{_{\mathrm{MAX}}}$. The behavior of this distance is, however, governed by the behavior of the sinusoidal function $\sin(\Delta\Phi_{_{\mathrm{GH}}}/2)$. When its argument is an integral multiple of $\pi$,
\begin{equation}
\Delta\Phi_{_{\mathrm{GH}}} \, = \, 2\,m\,\pi \,\,\,\,\,\,\,\,\,\,\,\,\,\,\,\,\,\,\,\, \mathrm{for} \,\,\,\,\,\,\,\,\,\,\,\,\,\,\,\,\,\,\,\, m \,=\, 0,1,2...
\end{equation}
the effect of the Goos-H\"anchen phase on the weak measurement is null and the usual formula for $\Delta Y_{_{\mathrm{MAX}}}$ is reconstructed. This occurs for particular combinations of $N$ and $\theta$, as can be seen in Fig.2, where the sinusoidal function is plotted for for various $N$ as a function of the incidence angle. We can see that there is a minimum number of internal reflections to trigger the reconstruction of the usual formula. For a borosilicate prism and for $N<8$, $\Delta\Phi_{_{\mathrm{GH}}}$ is never an integer multiple of $2\pi$. As the number of reflections increases, however, this result becomes accessible to more angles, starting with two angles for $N=8$.

Fig.3 shows the effect of the Goos-H\"{a}nchen phase on $\Delta Y_{_{\mathrm{MAX}}}$ for $N=8$ and $N=16$. The dashed line corresponds to the standard weak measurement formula,

\[\Delta Y_{_{\mathrm{MAX}}} \,=\, \Delta y_{_{\mathrm{GH}}}/|\Delta\epsilon|,\]
while the solid line represents the the weak measurement with phase. We can see that the Goos-H\"{a}nchen phase renders the distance between intensity peaks virtually zero, except for angles at which $\Delta\Phi_{_{\mathrm{GH}}}=2\,m\,\pi$. This effect is strongly local, with $\Delta Y_{_{\mathrm{MAX}}}$ falling rapidly to zero around such incidence angles. We can estimate how narrow the $\Delta Y_{_{\mathrm{MAX}}}$ peaks are by calculating the angles at which they fall to half their maximum value. Expanding the sinusoidal function up to the first order around the maximum, we have
\begin{eqnarray}\label{eq:hlf}
\frac{\displaystyle \Delta y_{_{\mathrm{GH}}}(\theta_{_{\mathrm{MAX}}})+\Delta y_{_{\mathrm{GH}}}'(\theta_{_{\mathrm{MAX}}})\,\delta }{\displaystyle |\Delta\epsilon|}
\frac{\displaystyle 1}{ 1+\left[\frac{\displaystyle \Delta\Phi_{_{\mathrm{GH}}}'(\theta_{_{\mathrm{MAX}}})}{\displaystyle 2\,|\Delta\epsilon|}\right]^{\2}\,\delta ^{\2}} \,=\, \frac{\displaystyle \Delta y_{_{\mathrm{GH}}}(\displaystyle \theta_{_{\mathrm{MAX}}})}{\displaystyle 2\,|\Delta\epsilon|}\,\,,
\end{eqnarray}
where
\begin{equation}
\delta \,=\, \theta_{_{1/2}}-\theta_{_{\mathrm{MAX}}}\,\,.
\end{equation}
From Eq.(\ref{eq:hlf}) we can find
\begin{eqnarray}{\label{eq:hlf2}}
\theta_{_{1/2}} \,=\, \theta_{_{\mathrm{MAX}}}+\left[\frac{\displaystyle 2\,|\Delta\epsilon|}{\displaystyle k\,\mathrm{w}(z_{_{\mathrm{CAM}}})\,\Delta y_{_{\mathrm{GH}}}}\right]^{\2}
\Bigg{\{}\frac{\displaystyle \Delta y_{_{\mathrm{GH}}}'}{\displaystyle \Delta y_{_{\mathrm{GH}}}}
\pm\sqrt{\left(\frac{\displaystyle \Delta y_{_{\mathrm{GH}}}'}{\displaystyle \Delta y_{_{\mathrm{GH}}}}\right)^{\2}+\left[\frac{k\,\mathrm{w}(z_{_{\mathrm{CAM}}})\,\Delta y_{_{\mathrm{GH}}}}{\displaystyle 2|\Delta\epsilon|}\right]^{\2}}\Bigg{\}}_{\theta_{_{\mathrm{MAX}}}}\,\,.
\end{eqnarray}

The angular distance between the half-peak intensity points for a borosilicate prism and $N=8$ is $0.819^{\circ}$ for the first peak and $1.299^{\circ}$ for the second, see Fig.3a. For comparison, for $N=16$, the first peak has an angular width of $0.032^{\circ}$ while its broadest peak, the third one, is only $0.648^{\circ}$ wide, see Fig.3b.

In Fig.4 we can see contour plots of $\Delta Y_{_{\mathrm{MAX}}}$ against $\Delta\epsilon$ and $\theta$. We can see that the greater the amplification sought, the smaller the angular range for the allowed incidence angles as well as smaller the value of the perturbation angle.

\section*{\shadowtext{V Conclusions}}

We have reexamined the optical weak measurement of Goos-H\"anchen shifts for a Gaussian beam transmitted through a dielectric block chain, considering the effect of the Goos-H\"anchen phase, and showing that it has a strong destructive influence on measurements. This result does not go against the literature on this subject, but simply addresses an overlooked point, that is, the effect of a relative Goos-Hänchen phase on Weak Measurements. In the excellent papers by Dennis and G\"otte describing the general theory of optical Weak Measurements for beam shifts, for instance \cite{OWM1, OWM2}, an operator approach is taken, but the particularities of the Fresnel coefficients in the Total Internal Reflection regime are not considered, while in the experimental work of Jayaswal et al. \cite{S2013} a system of waveplates is used to remove the relative Goos-H\"{a}nchen phase.

Our analysis provides an alternative way of removing the effect of this relative phase and re-obtain the standard theoretical formula of weak measurements. Since the distance between intensity peaks, which is used to calculate the relative Goos-H\"{a}nchen shift, depends on the relative phase through a sine function, and since the relative phase is a function of the incidence angle and of the number of internal reflections, we can control these parameters in order to null the sine function. As can be seen in Fig. 2, however, this is not always possible. For a borosilicate prism $(n=1.515)$, incidence angles for which the sine function is zero only appear for a number of internal reflections equal or greater than 8. Nevertheless, after this threshold, it is always possible to find such angles, and their number increases with the number of internal reflections.

In Fig.3 we can compare the standard optical weak measurement (dashed red line) with the weak measurement with phase (solid blue line). The relative Goos-H\"{a}nchen phase completely destroys the measurement, except around the angles for which $\sin\left(\Delta\Phi_{_{\mathrm{GH}}}/2\right)=0$, providing a way of having the standard result without the aid of waveplates. This method, however, demands accuracy in the selection of the incidence angle. Using Eq. (\ref{eq:hlf2}) we can calculate the angular width around the angle that maximizes the distance between intensity peaks for which such a distance is at least half of its maximum value. We can see that increasing the number of internal reflections the peaks become greater, but the application of this amplification has to be considered against the available precision in the choice of the incidence angle. For $N=8$ and $N=16$ there are peaks for $\theta=4.63^{\circ}$ and $\theta=15.37^{\circ}$, but their angular width is $0.819^{\circ}$ and $1.299^{\circ}$ for $N=8$ and $0.407^{\circ}$ and $0.648^{\circ}$ for $N=16$, respectively.

Finally, we notice that transverse shifts, such as the Imbert-Fedorov effect \cite{IFe}, can also be subject to a similar analysis since it is generated by polarization-dependent phase shifts in the Total Internal Reflection regime. Such an analysis cannot, however, be extended to angular deviations. Despite the fact that Weak Measurements have been successfully employed in the study of such phenomena \cite{WMAngGHExp}, the mechanism behind them is quite different than the mechanism behind lateral and transverse shifts. In both the Angular Goos-H\"anchen shift and in the Fresnel Filtering \cite{AngGHFF}, the angular shift is caused by the symmetry breaking of the incoming beam by the Fresnel coefficients \cite{SymBrk}. Since symmetry breaking requires real Fresnel coefficients, no additional phase is gained by the electric field, and the phase removal problem ceases to be a concern. Another system that is an interesting topic for future research is that of reflection on metallic surfaces, since the Fresnel coefficients for this case are complex, the reflected beam will always gain an additional phase.

We hope our results stimulate interest in the effects of relative phases on Weak Measurements on other phenomena such as the Imbert-Fedorov and the Optical Spin Hall effects \cite{OSHE}, high precision phase estimations \cite{S2013PRL}, the use of Mach–Zehnder interferometers to detect glucose concentration \cite{S2016} and the amplification of the time delays in quantum mechanics\cite{S2016Nat}, to give only a few examples.



\newpage

\WideFigure{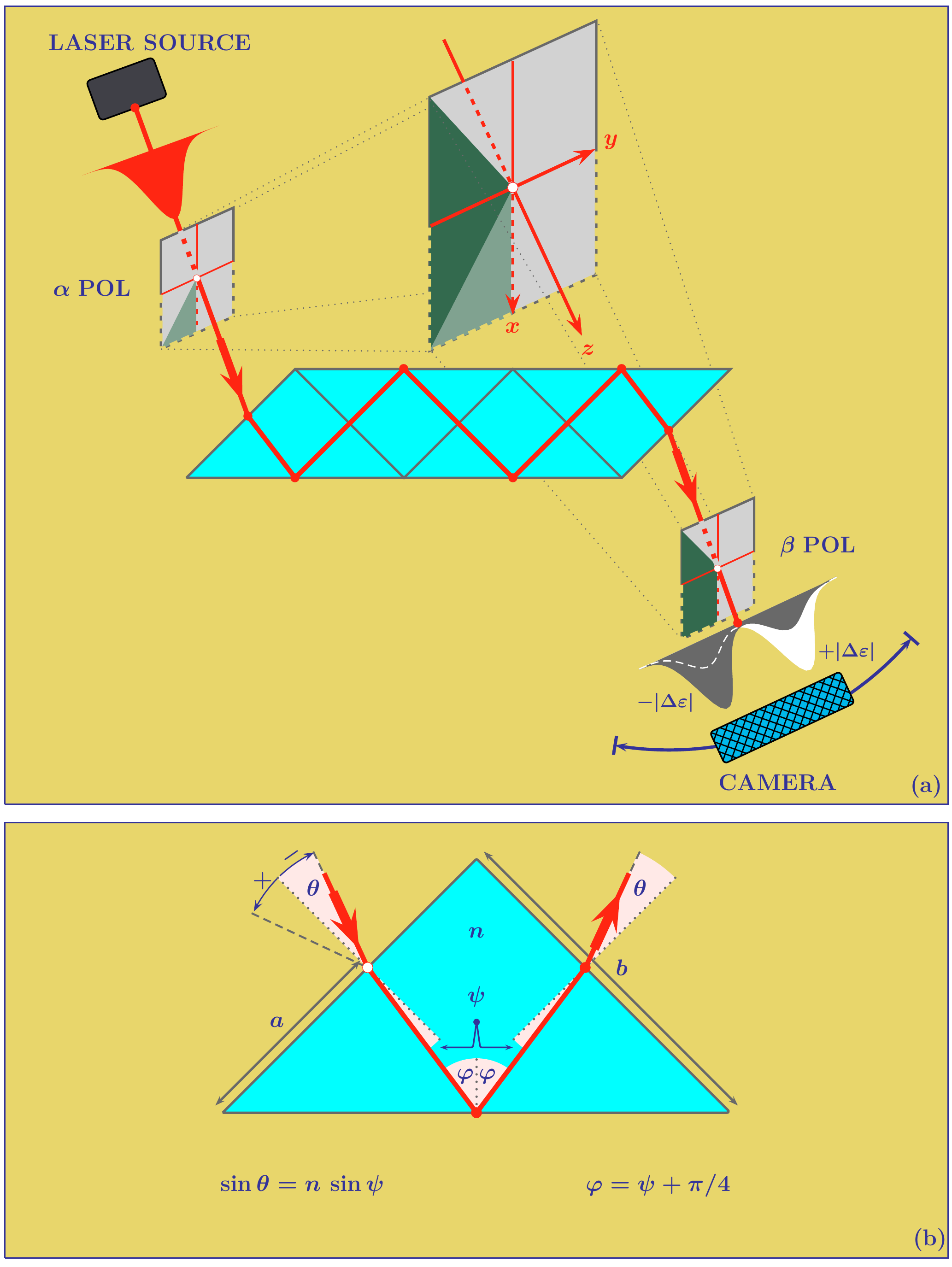}{
(a) The schematic representation of an optical weak measurement experimental set-up. The optical beam coming from the source passes through the $\alpha$ polarizer which makes it an equal mixture of TE and TM polarization states. The beam then enters the dielectric structure, in the total internal reflection regime, coming then out of it and passing through the $\beta$ polarizer. This polarizer mixes the two polarized components of the beam which, finally, arrives at camera. The intensity profile collected by the camera leading to the weak measurement amplification is controlled by a perturbation $\pm|\Delta\varepsilon|$) of the $\beta$ polarizer's angle.
(b) The basic building block of the dielectric structure: A right angle triangular prism with refraction index $n$. The relevant angles are also represented. $\theta$ and $\psi$ are the angles of the incident and refracted beams, respectively. They are related through the Snell's law. $\varphi$ is the internal reflection angle and its relation to the refraction angle is set by the geometry of the system.
}

\WideFigure{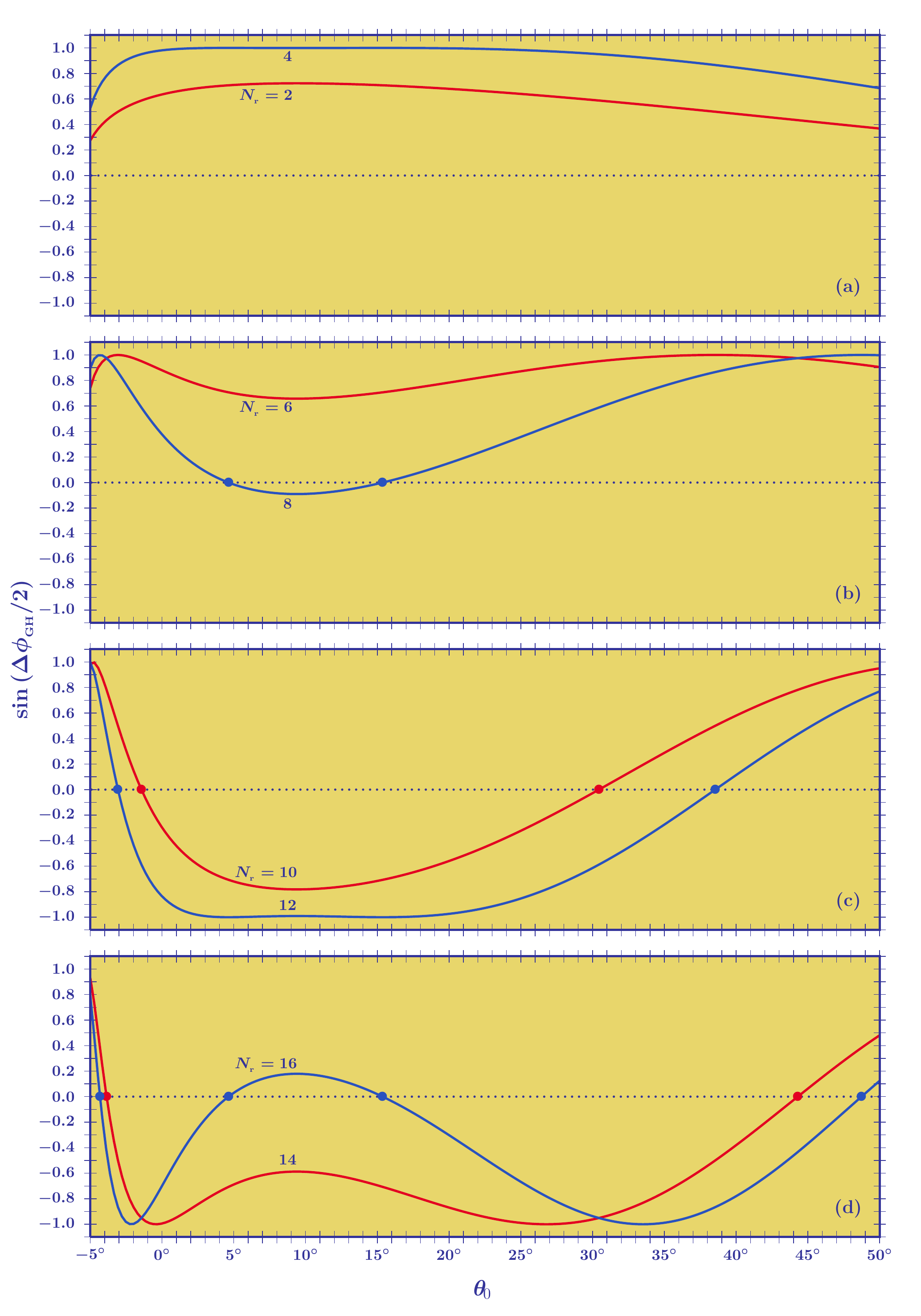}{$\sin\left(\Delta\Phi_{_{\mathrm{GH}}}/2\right)$ as function of the incidence angle for different values of $N$ with
$\mbox{w}_{\0}=1\,{\mathrm{mm}}$ and $n=1.515$ (borosilicate glass).}
\WideFigure{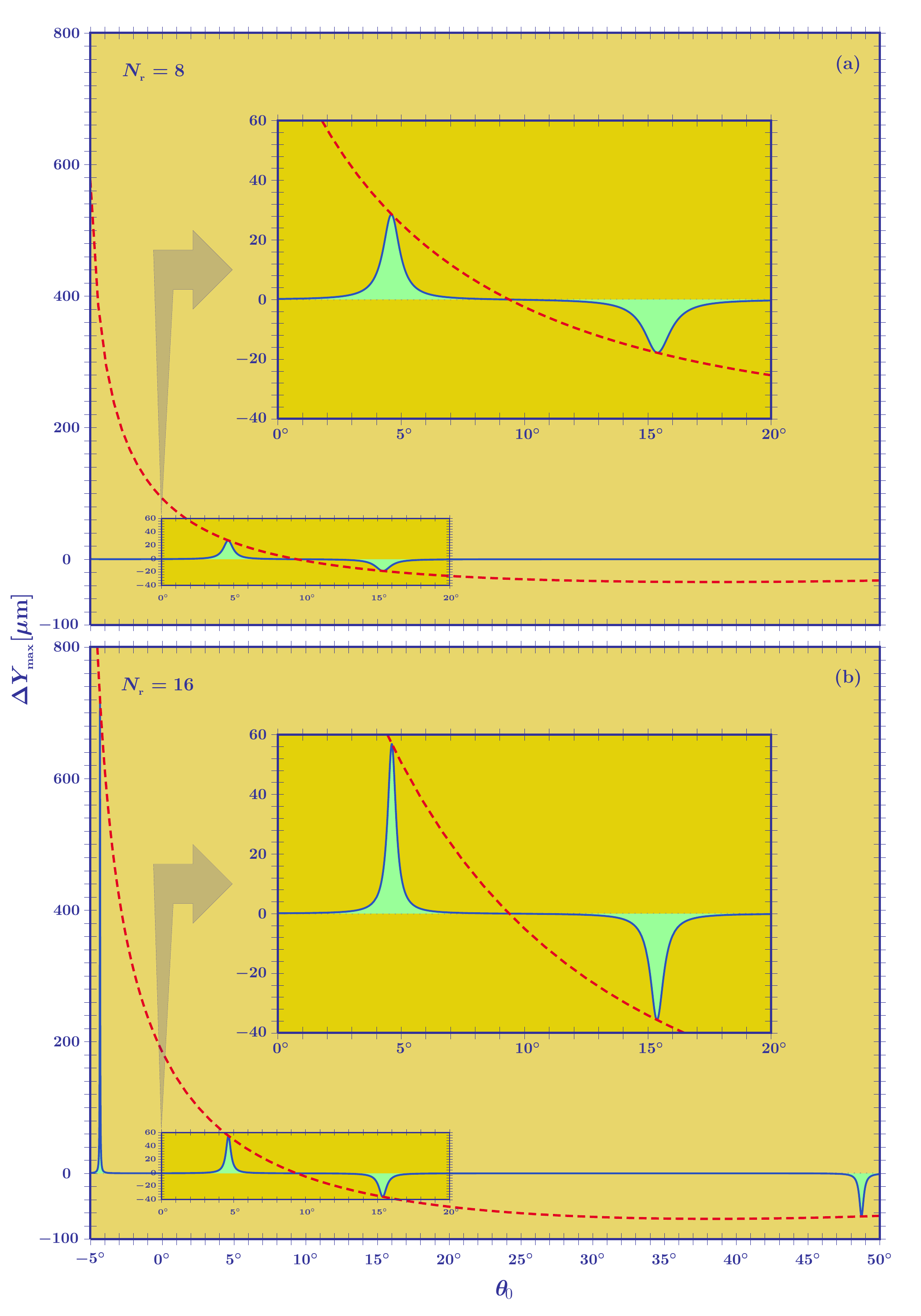}{Weak measurement amplification. The plot shows the distance between the intensity peaks as function of the incidence angle for $N=8$ and $16$ internal reflections, $\mbox{w}_{\0}=1\,{\mathrm{mm}}$,
$\lambda=633\,\mu{\mathrm{m}}$ and $n=1.515$ (solid blue line). The maximums are located where the sinusoidal function is null.
The dashed red line represents the curve of the standard formula in the literature.}
\WideFigure{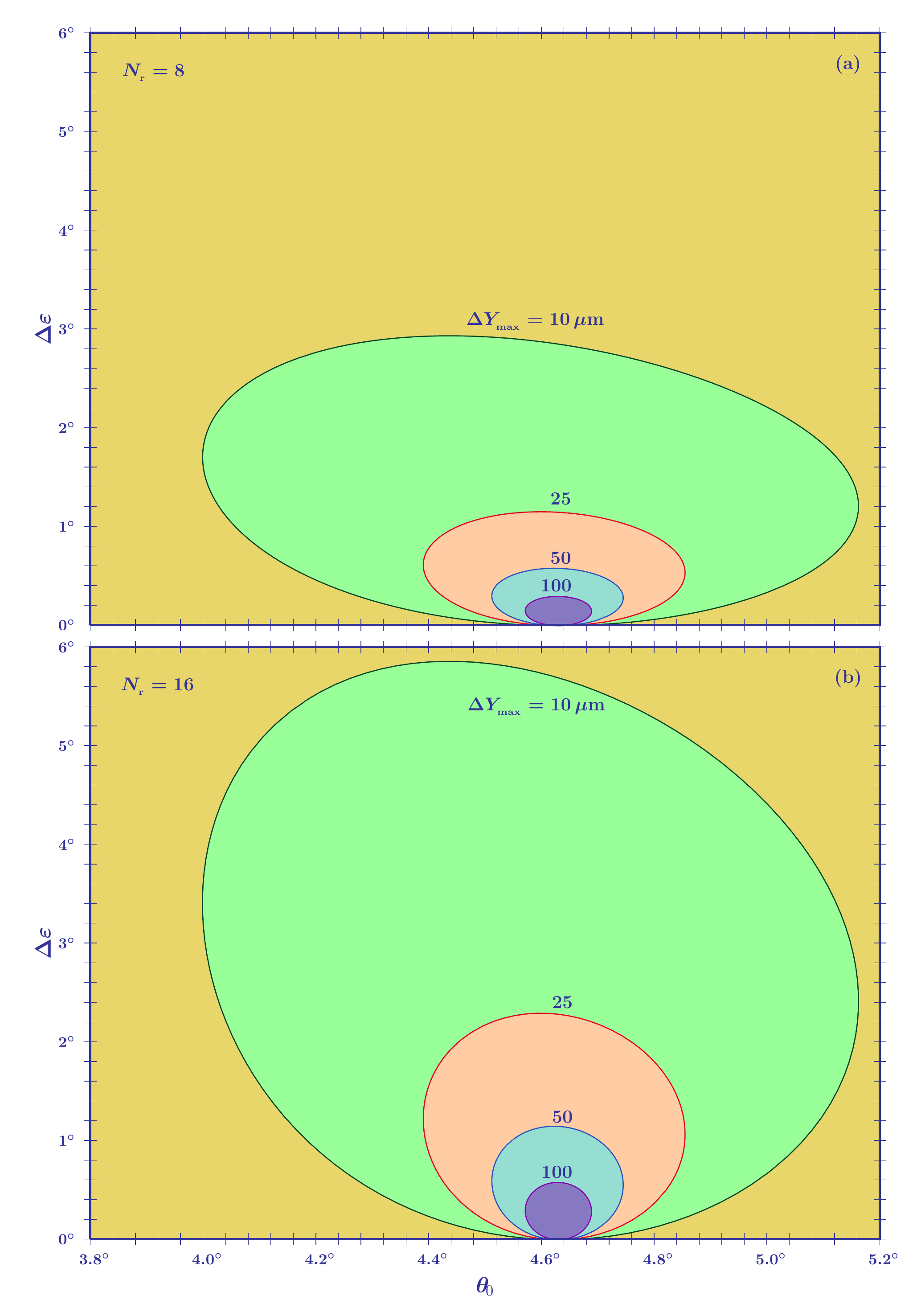}{Level curves of $\Delta Y_{_{\mathrm{MAX}}}$,
plotted as a function of the incidence angle $\theta_{\0}$ and the perturbation angle $\Delta\varepsilon$ of the second polarizer, for a beam waist of $\mbox{w}_{\0}$=1 mm and refraction index  $n=1.515$.}

\end{document}